\documentclass{aa}
\usepackage{graphicx,natbib}
\bibpunct{(}{)}{;}{a}{}{,}
\begin{document}

   \title{A Strongly Nonlinear Alfv\'enic Pulse in a Transversely Inhomogeneous
Medium}

   \author{D. Tsiklauri, V.M. Nakariakov and T.D. Arber}

   \offprints{David Tsiklauri, \\ \email{tsikd@astro.warwick.ac.uk}}

   \institute{Physics Department, University of Warwick, Coventry,
   CV4 7AL, England}

   \date{Received ???? 2002 / Accepted ???? 2002}

\abstract{ We investigate the interaction of a plane,
linearly polarized, Alfv\'enic pulse with a one-dimensional,
perpendicular to the magnetic field, plasma density inhomogeneity
in the strongly nonlinear regime. Our numerical study of the full
MHD equations shows that: 
(i)  Plasma density
inhomogeneity substantially enhances
(by about a factor of 2) the generation of longitudinal
compressive waves. 
(ii) Attained maximal values of the
generated transverse compressive perturbations are  insensitive to
the strength of the plasma density inhomogeneity, plasma $\beta$ 
and the initial amplitude of the Alfv\'en wave.
Typically, they reach about 40\% of the initial Alfv\'en 
wave amplitude. 
(iii) Attained
maximal values of the generated relative density perturbations are
within  20-40\% for $0.5 \leq \beta \leq 2.0$. They depend upon
plasma $\beta$ strongly; and scale almost linearly
with the initial Alfv\'en wave amplitude. 
\keywords{Magnetohydrodynamics(MHD)-- waves -- Sun: activity --
Sun: Solar wind} }

\titlerunning{A Strongly Nonlinear Alfv\'enic Pulse ...}
\authorrunning{Tsiklauri, Nakariakov, \& Arber  }
\maketitle

\section{Introduction}

The Alfv\'en waves are usual candidates for energy transport from
the lower layers of the solar atmosphere to the corona, e.g.
\citep{mg,br}. However, efficient deposition of the
momentum and energy require interaction of linearly incompressible
Alfv\'en waves with compressible magnetoacoustic waves, e.g.
\citep{od2,od3,ons00}. Also,
the compressible waves, in contrast to the Alfv\'en waves, can
transport energy and momentum across the magnetic field, spreading
out the heated region. In addition, observational detection of
Alfv\'en waves in open structures of the corona can be based upon
measurement of the compressible fluctuations, e.g.
\citep{of97,of98,of00}. These are generated in the lower corona by the Alfv\'en
waves through linear or nonlinear mechanisms, e.g. \citep{noa}.
This method can be complimentary to the observation
of coronal Alfv\'en waves through non-thermal broadening of
emission lines, e.g. \citep{btdw98}.

In the inner heliospheric solar wind, Alfv\'en waves are observed
{\it in situ} and represent the main component in MHD turbulence
\citep{tm,th}. This suggests another interesting problem: why the
incompressible turbulence dominates in the solar wind, and why
compressible fluctuations are not observed, despite the
theoretical possibility for these two kinds of MHD fluctuations to
be coupled because of the medium inhomogeneity and nonlinearity.
Thus, the study of coupling of compressible and incompressible
fluctuations is important to the physics of the solar corona and
the solar wind. There are several possible mechanisms for the
coupling.

The decay instability of Alfv\'en waves is one of the possible
examples of interaction between the MHD wave modes. This mechanism
involves {\it resonant} three-wave interaction of Alfv\'en and
magnetoacoustic waves (e.g. \citet{sg} and references therein). The
efficiency of this interaction is governed by the amplitudes of
the interacting waves. However, this mechanism works only for
quasi-periodic (perhaps wide-spectrum, \citet{malara2000}) waves,
and is not efficient for the wave pulses that could be generated
by some transient events on the Sun such as solar flares and
coronal mass ejections, e.g. \citep{r01}.

In contrast, the efficiency of {\it non-resonant} mechanisms of
the compressible fluctuation excitation by the Alfv\'en waves does
not depend on the coherentness of Alfv\'en perturbation and
consequently work even for single wave, wide-spectrum, pulses.

Nonlinear excitation of magnetoacoustic perturbations by nonlinear
elliptically polarized Alfv\'en waves via the longitudinal
gradients of the total pressure perturbations is one of the
possible examples of the non-resonant MHD wave interaction. In
this mechanism the generation of compressible perturbations
results in the self-interaction and subsequent steepening of the
Alfv\'en wave front, which is described by the Cohen-Kulsrud
equation, e.g. \citep{ck,vnl,noa}. In the following, we refer to
this mechanism as \lq\lq longitudinal", to distinguish it from
the \lq\lq transverse mechanism", which generates compressible
perturbations via the transverse gradients of the total pressure
perturbations in nonlinear Alfv\'en waves. More detailed
discussion of these two mechanisms is presented in
\citet{nrm97,nrm98,Botha,td1}.

The transverse mechanism is dramatically modified in the case when
MHD waves interact with transverse stationary inhomogeneity of the
plasma. If the Alfv\'en speed is {\it inhomogeneous} across the
magnetic field, initially plane, linearly polarized Alfv\'en waves
become oblique and sharp gradients in the direction across the
field are secularly generated. This constitutes a basis of the
well known Alfv\'en wave phase mixing phenomenon which has been
extensively investigated in connection with the solar coronal
heating with MHD waves \citep{awpm}. Various aspects of this
phenomenon have been intensively studied using full MHD numerical
simulations \citep{malara,od1,od2,ptbg,glb,iha}. As demonstrated
by \citet{nrm97,nrm98,Botha,td1}, phase mixing of Alfv\'en waves in
the {\it compressible} plasma, in the {\it weakly} nonlinear
regime leads to the generation of fast magnetoacoustic waves, and
various regimes of this process, relevant to solar coronal and
heliospheric applications have been studied. In particular, it has
been found that the inhomogeneity of the plasma {\it across} the
magnetic field, associated with various types of structuring (e.g.
plumes in the coronal holes, boundaries between the slow and the
fast solar winds, flow tubes, etc.), plays the crucial role in the
interaction of compressible and incompressible weakly-nonlinear
MHD modes.

Problems connected with the interpretation of MHD fluctuations
observed in the solar wind and the propagation of intensive MHD
wave pulses in the solar corona (e.g., flare and CME-associated
waves) require detailed study of the interaction of the MHD
pulses with plasma inhomogeneities in the {\it strongly} nonlinear
regime. Our interest to this regime is motivated by the necessity
to account for effects of higher order nonlinearities and the
back-reaction of nonlinearly generated compressible perturbations
on the source Alfv\'en pulse. Also, as the efficiency of the
\lq\lq transverse" mechanism is connected to the longitudinal
wave number (together with the transverse wave number and the
amplitude), and as the \lq\lq longitudinal" mechanism increases,
through wave steepening,
the longitudinal wave numbers, we expect that the simultaneous
action of these two mechanisms can enhance the efficiency of the
transverse mechanism. In this work, we study, by direct numerical
simulations, the generation of compressible fluctuations by a
strongly nonlinear Alfv\'enic pulse interacting with a transverse
plasma inhomogeneity.

The paper is organized as follows: in Section~2 we describe our
model and the numerical method applied, in Section~3 the results
of the simulations are discussed separately in the high and low
$\beta$ cases (subsections 3.1 and 3.2 respectively), sub-section
3.3 deals with the investigation of parametric space, and finally,
the conclusions are presented in Section~4.

\section{The model}

The model studied here is similar to one discussed in \cite{td1}:
we use the equations of ideal MHD
$$
\rho {{\partial \vec V}\over{\partial t}} +
\rho(\vec V \cdot \nabla) \vec V = - \nabla p -{{1}\over{4 \pi}}
\vec B \times {\rm curl} \vec B, \eqno(1)
$$
$$
{{\partial \vec B}\over{\partial t}}= {\rm curl} (\vec V \times \vec B),
\eqno(2)
$$
$$
{{\partial p}\over{\partial t}} + \vec V \cdot \nabla p + \gamma p
\nabla \cdot \vec V=0,
\eqno(3)
$$
$$
{{\partial \rho}\over{\partial t}} + {\rm div}(\rho \vec V)=0,
\eqno(4)
$$
where $\vec B$ is the magnetic field, $\vec V$ is plasma velocity,
$\rho$ is plasma mass density, and $p$ is plasma thermal pressure
for which the adiabatic variation law is assumed.

We solve equations (1)--(4) in Cartesian coordinates ($x,y,z$) and
under the assumption that there is no variation of the physical
values in the $y$-direction, i.e. ($\partial / \partial y =0$,
2.5D approximation) with the use of {\it Lare2d} \citep{Arber}.
{\it Lare2d} is a numerical code which operates by taking a
Lagrangian predictor-corrector time step and after each Lagrangian
step all variables are conservatively re-mapped back onto the
original Eulerian grid using Van Leer gradient limiters. This code
was also used to produce the results in Refs. \citet{Botha,td1}. As
in  \citet{Botha,td1}, the equilibrium state is taken to be
a uniform magnetic field $B_0$ in the $z$-direction and an
inhomogeneous plasma of density $\rho_0(x)$,
$$
\rho_0(x)=\rho_* \left(3-2 \, \tanh(\lambda x) \right). \eqno(5)
$$
Here, $\lambda$ is a free parameter which controls the steepness
of the density profile gradient.  The latter is localized
around $x=0$. The temperature profile $T_0(x)$ is set up to allow
for the total pressure to be constant everywhere. In our
normalization, which is the same as that of \citet{Botha,td1},
${\vec B}=B_0 {\bar {\vec B}}$, ${\vec r}=a_* {\bar {\vec r}}$,
$C_A(x)=B_0 / \sqrt{4 \pi \rho_0(x)}= \left[B_0 / \sqrt{4 \pi
\rho_*}\right]/\sqrt{3-2 \, \tanh(\lambda x)}= C^*_A/ \sqrt{3-2 \,
\tanh(\lambda x)}=C^*_A {\bar C_A(x)}$, $t=(a_*/C^*_A){\bar t}$, ${\vec
V}=C^*_A {\bar {\vec V}}$, the dimensionless local Alfv\'en speed
is ${\bar C_A(x)}=1/ \sqrt{3-2 \, \tanh(\lambda x)}$. $a_*$ and $\rho_*$
are the units of length and density respectively. In what follows
we omit bars on all dimensionless physical quantities.

\section{Numerical Results}

We set up the code in such a way that initially longitudinal
($V_z$) and transverse ($V_x$,$B_x$,$B_z$) perturbations and
the density perturbation, $\rho$, (perturbed by both
longitudinal and transverse modes) are absent and the  initial
amplitude of the Alfv\'en pulse is strongly non-linear, i.e.
typically A=0.5. At $t=0$, the Alfv\'en perturbation is a plane
(with respect to $x$-coordinate) pulse, which has a Gaussian
structure in the $z$-coordinate,
$$
B_y(z)=A \exp \left(-{{z^2}\over{\delta}}\right),\ 
V_y(x,z)=-C_A(x)\, B_y(z). \eqno(6)
$$
Here, $\delta$ is a free parameter which controls the width of the
initial pulse. In all our numerical runs $\delta=0.1$.

As it is discussed in Introduction  (cf. \citet{nrm97} for
details) in the considered geometry, Alfv\'en waves represented by
$V_y$ and $B_y$, are incompressible up to the {\it cubic} nonlinearity. 
The longitudinal (represented by $V_z$) and the transverse
(represented by $V_x$,$B_x$,$B_z$) compressible perturbations
(of course, both perturbing $\rho$) are generated nonlinearly by the
longitudinal and transverse gradients of the total pressure.
The efficiency of the generation depends upon the amplitude
of the Alfv\'en waves, in other words, {\it the incompressible and
compressible perturbations are  linearly decoupled}.
The latter guarantees that with the choice of our initial
conditions the compressible perturbations are indeed initially absent
from the system, which {\it a priori} is not clear if the
linear coupling is present (cf. \citet{td2}).
In addition, the transverse compressible perturbations
are generated by plane Alfv\'en waves only in the 
presence of the transverse profile of the local Alfv\'en speed.

The simulation box size is set by the limits $-15.0 < x < 15.0$
and  $-15.0 < z < 15.0$. The pulse starts to move from
point $z=-12.5$ towards the positive $z$'s.
Both the scale of density inhomogeneity and width of the
Alfv\'enic pulse are much smaller than the size of the
calculation domain.

We have performed calculations on various resolutions in an
attempt to achieve convergence of the results. The graphical
results presented here are for the spatial resolution $2500 \times
2500$, which refers to number of grid points in $z$ and $x$
directions respectively. We have used a non-uniform grid in our
simulations, namely, in $x$ direction 75\% of the grid points
where concentrated between $-5.0 \leq x \leq 5.0$ where  the
spatial inhomogeneity has strongest gradients. We have also
performed calculation on the spatial resolution $3500 \times 3500$
and found that the maximal generation levels (maximum of an
absolute value over the whole simulation domain) for all physical
quantities as a function of time are the same as in the case of
$2500 \times 2500$ resolution. Thus, the results presented here
are, indeed, converged.

Our main numerical runs are split in two parts, treating separate
cases when plasma $\beta$, which is the ratio of speed of sound to
the Alfv\'en velocity squared, is less and greater than unity.
This split is motivated by weakly nonlinear results, showing that
the particular scenario of the interaction of MHD waves is
determined by the ratio $(C_s/C_A)$ being less or greater then
unity, e.g. \citep{ck}.

\subsection{Case when $\beta > 1$}

In this subsection we present solution of the Eqs.~(1)-(4) with
the above described equilibrium and the initial conditions for the
case when plasma-$\beta$ is 2.0. Here, $\lambda$-parameter was
fixed at 0.75.

\begin{figure}[]
\caption{Top panel: snapshot of $V_x/C_A(x)$ at $t=15.0$.
Bottom panel: contour-plot of $V_x/C_A(x)$ at the same instance.
Here, initial amplitude, $A=0.5$, density inhomogeneity steepness,
$\lambda=0.75$, plasma $\beta=2$.}
\end{figure}

Figures~1--3 show snapshots of the initially absent transverse
($V_x$) and longitudinal ($V_z$) components of the velocity
(representing the \lq\lq transverse" and \lq\lq longitudinal"
compressible waves, respectively) and density perturbation, at
time $t=15$.

\begin{figure}[]
\caption{Top panel: snapshot of $V_z/C_A(x)$ at $t=15.0$.
Bottom panel: contour-plot of $V_z/C_A(x)$ at the same instance.
Here, the parameters are the same is in Fig.~1. }
\end{figure}
The transverse compressible perturbations ($V_x$) are generated by
the transverse gradients of the total pressure perturbations.
In turn, these perturbations are generated by the Alfv\'en wave
phase mixing, and are located near $x=0$ where the density gradients
are the largest. Then they propagate across the field (see Fig.~1). The
longitudinal compressive perturbations, $V_z$ are generated even
in the absence of the plasma density inhomogeneity (Fig.~2). In
the contour plot (Fig.~2, bottom panel) it can be seen that there
are two wave fronts: one moving at a local Alfv\'en speed, and
another (compressive one) that moves faster than the local
Alfv\'en speed because $\beta=2$. Fig.~3 shows the relative
density perturbation associated with the compressible waves that
have been generated.
\begin{figure}[]
\caption{Top panel: snapshot of $(\rho-\rho_0)/ \rho_0$ at $t=15.0$.
Bottom panel: contour-plot of $(\rho-\rho_0)/ \rho_0$ at the same instance.
Here, the parameters are the same is in Fig.~1. }
\end{figure}

In Fig.~4, the top panel  presents spatial variation (across the
$x$-coordinate) and the dynamics of the transverse and
longitudinal compressible waves, and relative density
perturbations produced by them. 
In what follows $max$ and $min$ refer to the maximum and 
minimum over the whole simulation box (i.e. space) of a dimensionless 
physical quantity at a given time instance, respectively.
These are fairly good and simple (scalar) quantities describing
the generation and/or decay of a 2.5D physical quantity,
dynamics of which is otherwise not so straightforward to comprehend.
There are five interesting
observations: (i) the decay of the $max(|V_y|/C_A(x))$ (on expense
of which the transverse ($V_x$) and longitudinal ($V_z$)
compressive  waves and associated density perturbation are
generated) occurs in the  middle -- where the plasma
inhomogeneity is the strongest. This demonstrates the importance
of the inhomogeneity. (ii) the right wing ($x>0$) decays faster
than the left ($x<0$) one. For the value of plasma $\beta$ used,
it is expected that shock dissipation would be greater where the
local Alfv\'en velocity is greater (in this case the right wing).
Note that when referring to shock dissipation we mean
artificial dissipation that guarantees proper shock capturing,
while at all times we remain in the framework of an ideal MHD (no
bulk dissipation included). Thus, the shock viscosity ensures that
we recover the weak solution to ideal MHD. (iii) the relative
density perturbation and longitudinal compressive wave
$max(|V_z|/C_A(x))$ are generated in the homogeneous parts of the
domain ($x \leq -5$, $x \geq 5$) too. However, in the middle of
the domain, where the inhomogeneity is strong we see that
inhomogeneity of plasma density enhances the generation of these
quantities by about a factor of 2. (iv) transverse compressive
wave $max(|V_x|/C_A(x))$, which is not generated in the absence of
density inhomogeneity, is generated in the middle of the domain.
(v) $max(|V_y|/C_A(x))$, on expense of which the transverse
($max(|V_x|/C_A(x))$) and longitudinal ($max(|V_z|/C_A(x))$)
compressive  waves are generated, has two dips in those places
where the other physical quantities have two bumps, which clearly
demonstrates the correct energy balance as well as the importance of
the inhomogeneity (where actually the bumps occur).

\begin{figure}[]
\resizebox{\hsize}{!}{\includegraphics{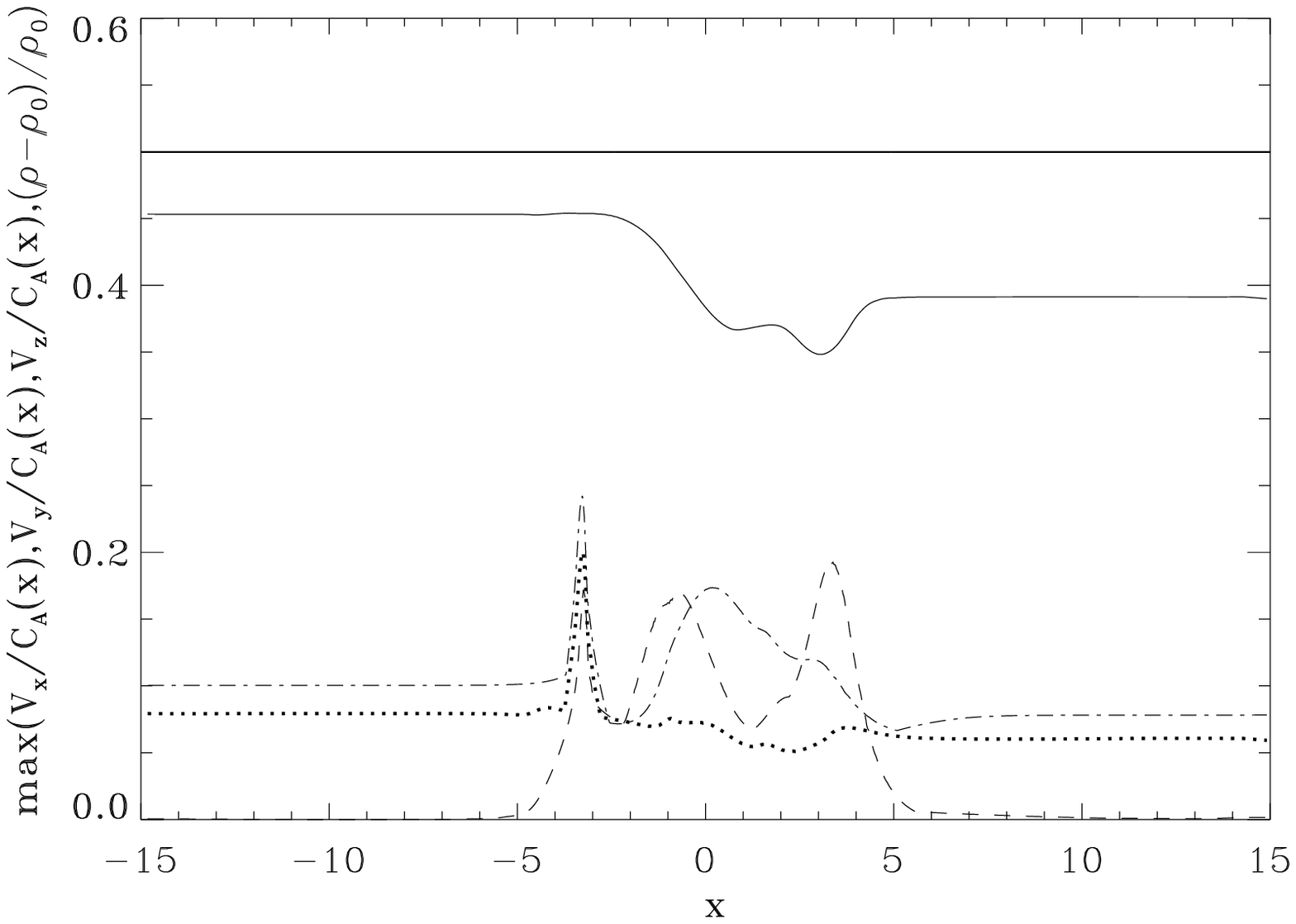}}
\resizebox{\hsize}{!}{\includegraphics{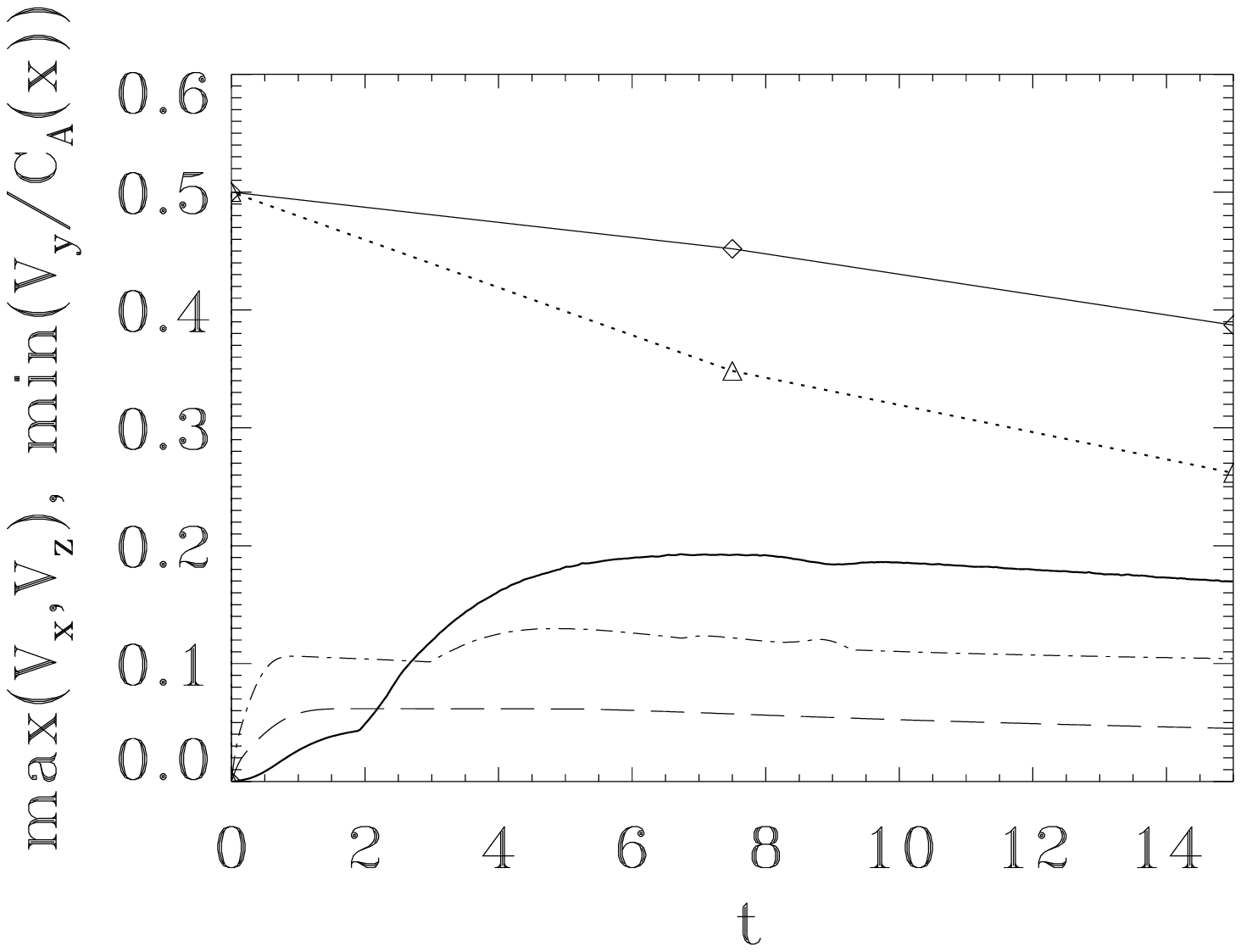}} \caption{Top
panel: spatial variation (across $x$-coordinate) and the dynamics
of non-linear generation of the transverse and longitudinal
compressive  waves as well as relative density perturbation in
time. Thick solid line corresponds to the initial value (at $t=0$)
of $max(|V_y|/C_A(x))$. Thin solid curve represents the same, but
for  $t=7.5$. The relative density perturbation
$max(|(\rho-\rho_0)/ \rho_0|)$ (dotted curve), the longitudinal
compressive wave $max(|V_z|/C_A(x))$ (dash-dotted curve), and the
transverse compressive wave $max(|V_x|/C_A(x))$ (dashed curve) are
given for $t=7.5$. Bottom panel: Evolution of $max(|V_x(x,z,t)|)$,
$min(|V_y(x,z,t)|/C_A(x))$, $max(|V_z(x,z,t)|)$ in time. The solid
curve with open rectangles presents decay of the initial Alfv\'en
perturbation ($min(|V_y(x,z,t)|/C_A(x))$) due to shock dissipation
in the case of absence of the plasma density inhomogeneity
($\lambda=0$) for three time instances ($t=0,7.5,15.0$). The
dotted curve  with open triangles depicts the same, but when
$\lambda=0.75$. The dash-dotted curve  represents
$max(|V_z(x,z,t)|)$, while thick solid curve corresponds to
$max(|V_x(x,z,t)|)$ both for the case of $\lambda=0.75$. The long
dashed curve represents $max(|V_z(x,z,t)|)$ for the case when
$\lambda=0$. Here, the parameters are the same is in Fig.~1.}
\end{figure}

In Fig.~4, the bottom panel  presents temporal variation of
transverse and longitudinal compressible wave amplitudes. In the
presence of the inhomogeneity ($\lambda=0.75$), the Alfv\'en
perturbation ($min(|V_y(x,z,t)|/C_A(x))$) decays much faster than
in the homogeneous case ($\lambda=0$).  In the inhomogeneous case,
the energy initially stored in the Alfv\'en wave in addition to
shock dissipation goes into the generation of the compressive
waves. The maximum amplitudes of longitudinal $max(|V_z(x,z,t)|)$
and transverse $max(|V_x(x,z,t)|)$ compressible waves attain a
substantial fraction of the initial Alfv\'en wave amplitude. In
the inhomogeneous case ($\lambda=0.75$), the longitudinal wave
attains about {\it twice} the maximal value than in the
homogeneous plasma case. Obviously, when $\lambda=0$ there is no
generation of the transverse compressive wave and
$max(|V_x(x,z,t)|)$ is identically zero for all times.

\subsection{Case when $\beta < 1$}

In this subsection we present solution of the Eqs.~(1)-(4) with
the above described equilibrium and the initial conditions for the
case when plasma-$\beta$ is 0.5.
\begin{figure}[]
\caption{Top panel: snapshot of $V_x/C_A(x)$ at $t=15.0$.
Bottom panel: contour-plot of $V_x/C_A(x)$ at the same instance.
Here, initial amplitude, $A=0.5$, density inhomogeneity steepness,
$\lambda=0.75$, plasma $\beta=0.5$.}
\end{figure}
As in the previous subsection,
$\lambda$-parameter was fixed at 0.75.
\begin{figure}[]
\caption{Top panel: snapshot of $V_z/C_A(x)$ at $t=15.0$.
Bottom panel: contour-plot of $V_z/C_A(x)$ at the same instance.
Here, the parameters are the same is in Fig.~5. }
\end{figure}

\begin{figure}[]
\caption{Top panel: snapshot of $(\rho-\rho_0)/ \rho_0$ at $t=15.0$.
Bottom panel: contour plot of $(\rho-\rho_0)/ \rho_0$ at
the same instance.
Here, the parameters are the same is in Fig.~5.  }
\end{figure}
\begin{figure}[]
\resizebox{\hsize}{!}{\includegraphics{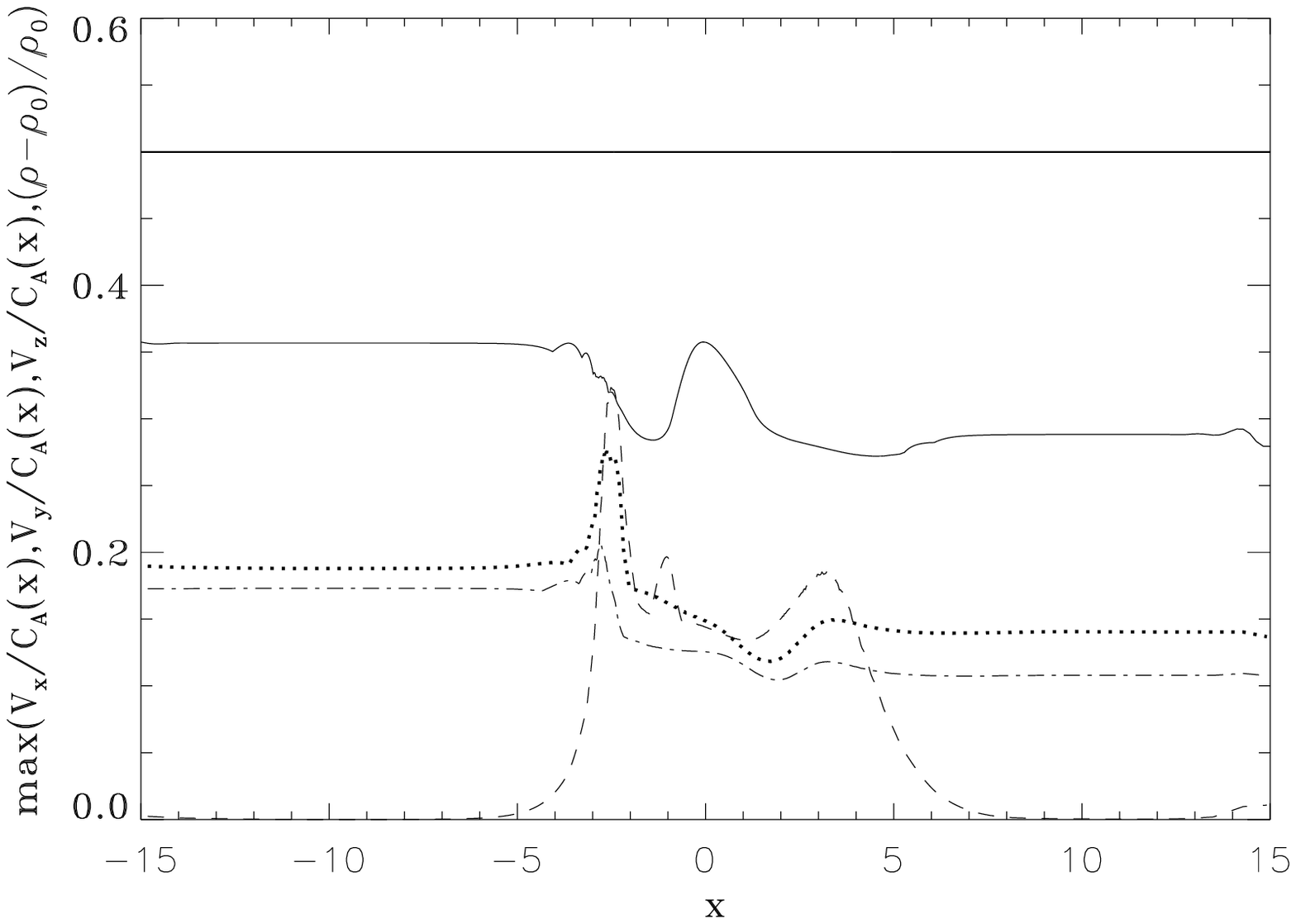}}
\resizebox{\hsize}{!}{\includegraphics{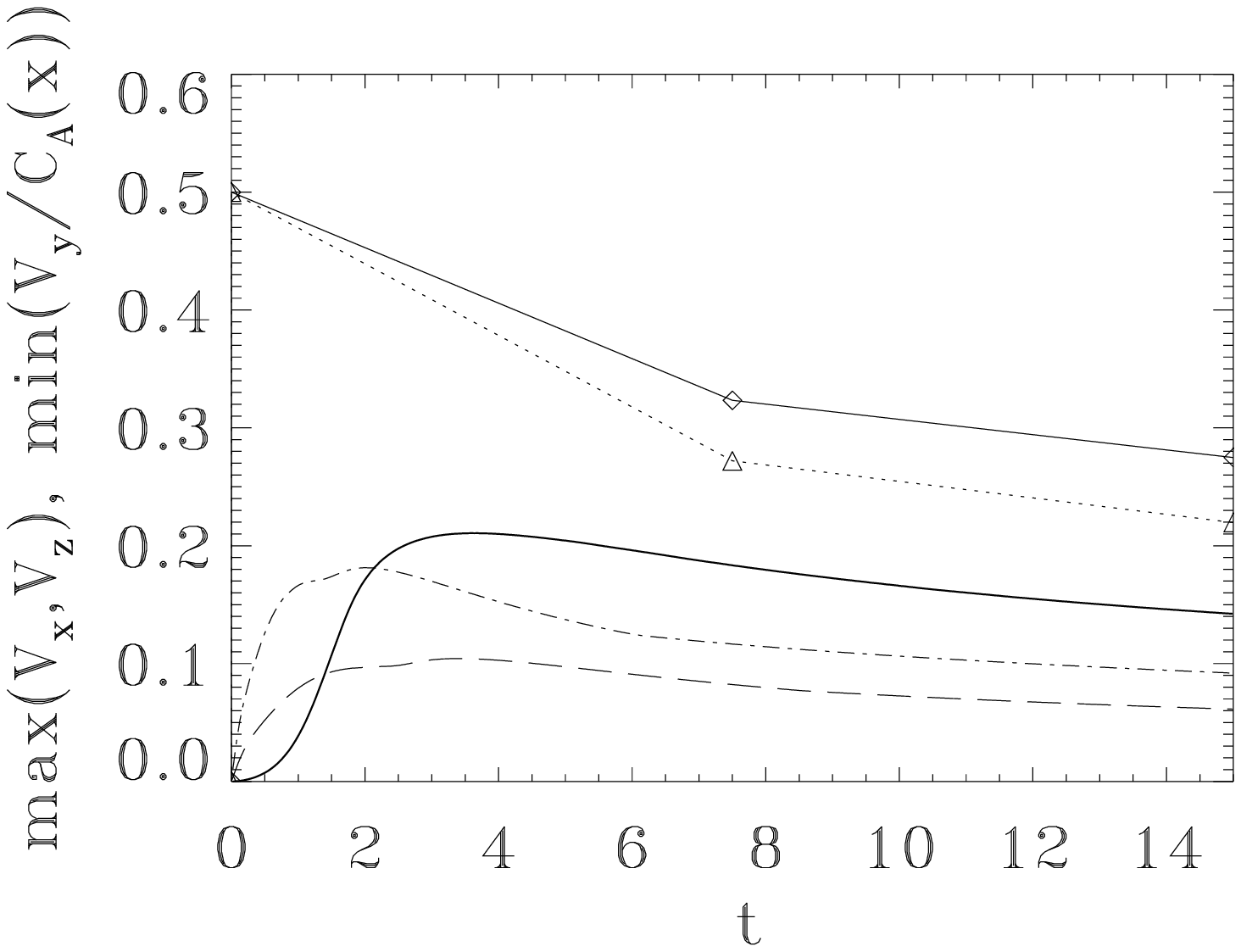}}
\caption{The same as in Fig. 4, but with  $\beta=0.5$.}
\end{figure}

Again, the initially absent transverse ($V_x$) and longitudinal
($V_z$) compressive waves and density perturbations are
efficiently generated (Figures~5--7) and reach a substantial
fraction of the initial Alfv\'en wave amplitude.

In the contour plots in Figs.~6 and 7
it can be seen that there are two wave fronts:
one moving at a local Alfv\'en speed, and another (compressive one)
that moves slower than
the local Alfv\'en speed because $\beta=0.5$ (compare with Figs.~2 and 3).
Therefore, since the velocity difference between the
first and  second parts of the solution is not as
great as in the $\beta=2$ case, these two parts
seem be blended into each other.

Figure~8 presents the evolution of transverse and longitudinal
compressive wave amplitudes in time. This graph is quite similar
to Fig.~4. The noteworthy difference is as follows: for the value
of plasma $\beta$ used here ($\beta=0.5$), the non-linear term in
the scalar Cohen-Kulsrud equation is larger than in the former
case of $\beta=2$ (Fig.~4). Therefore in Fig.~8, top panel, we
observe that the shock dissipation of the Alfv\'en wave is greater
(the thin solid line goes further down than in Fig.~4, also the
top panel), and, in turn, the non-linear generation of the
transverse and longitudinal compressive is further enhanced.

\subsection{Parametric study}

In this subsection we explore the parametric space of the
problem. In particular, we investigate how the
maximal value of the generated transverse compressive wave
depends on the plasma density inhomogeneity parameter, $\lambda$,
plasma $\beta$, and initial amplitude of the Alfv\'en wave, $A$.

In Fig.~9A we plot the dependence of the maximum of the absolute
value of the transverse compressive perturbation,
$max(|V_x(x,z,t)|)/A$, versus $\lambda$ for $\beta=2$ (solid
curve) and $\beta=0.5$ (dashed curve). There are two noteworthy
features in this graph, first, the maximal value of the generated
transverse compressive wave depends on the plasma density
inhomogeneity parameter rather weakly (once $\lambda \geq 0.4$),
and second, efficiency of the generation of $V_x$ is somewhat
larger in the $\beta=0.5$ case than in the case of $\beta=2$.
There are no data points between $0 \leq \lambda \leq 0.4$ because
in order to accommodate the inhomogeneity in the simulation domain
we would have to increase its size, which was not possible with
available computational resources. Also, we made 5 runs of {\it
Lare2d} code for the different values of $\beta$. 
The results are presented in Fig.~9B. We gather
from the graph that the maximal value of the generated transverse
compressive wave depends on the plasma $\beta$ rather weakly.

\begin{figure*}
\centering
 \includegraphics[width=17cm]{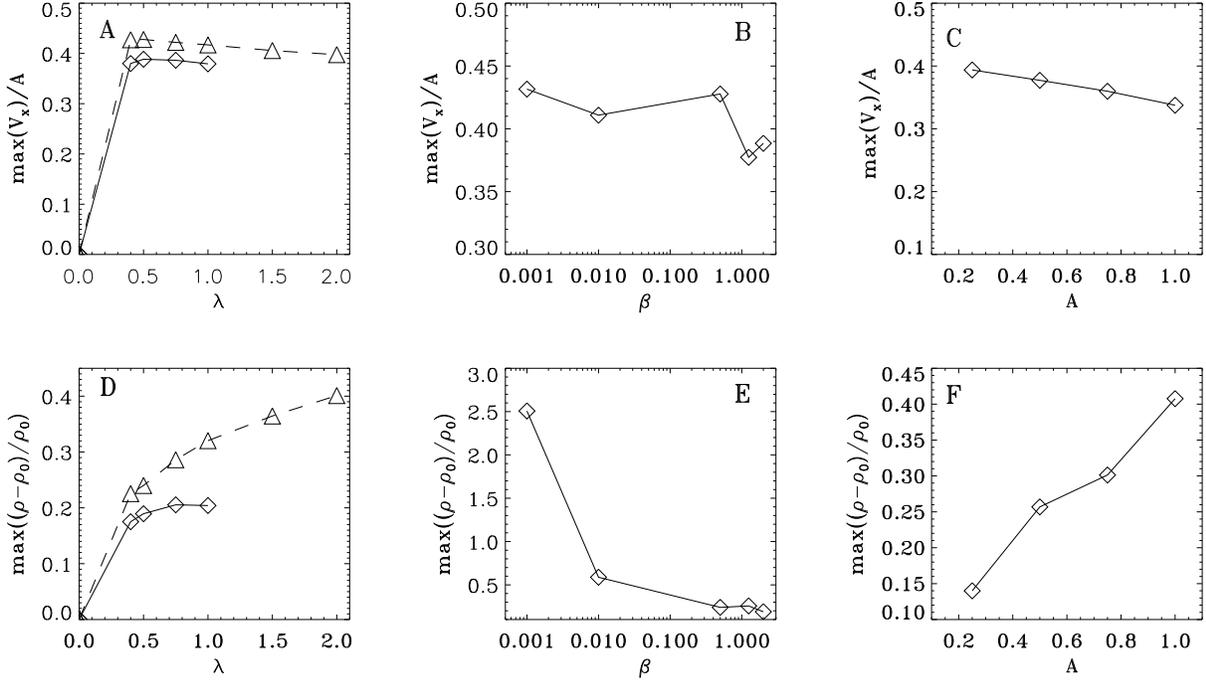}
\caption{(A) Dependence of $max(|V_x(x,z,t)|)/A$
versus $\lambda$ (density inhomogeneity steepness) for
$\beta=2$ (solid curve) and $\beta=0.5$
(dashed curve). Here, initial amplitude $A=0.5$.
(B) The same, but
versus $\beta$ for $\lambda=0.5$ and $A=0.5$. 
(C) The same, but
versus $A$ for $\lambda=0.5$ and $\beta=1.25$.
(D) Dependence of $max(|(\rho-\rho_0)/ \rho_0|)$
versus $\lambda$ for
$\beta=2$ (solid curve) and $\beta=0.5$
(dashed curve) both for $A=0.5$.
(E) The same, but
versus $\beta$ for $\lambda=0.5$ and $A=0.5$.
(F) The same, but
versus  $A$ for $\lambda=0.5$ and $\beta=1.25$. } \label{fig9_f}
\end{figure*}

Yet another valuable insight can be obtained by studying
the dependence of the maximal value of  generated transverse
compressive wave on the initial amplitude of the Alfv\'en wave
as our problem is essentially non-linear.
In Fig.~9C we plot results of  numerical runs for different
values of $A$, while $\lambda$ and $\beta$ where fixed at
0.5 and 1.25 respectively.
We gather from Fig.~9C that quite unexpectedly the ratio
$max(|V_x(x,z,t)|)/A$ is  insensitive to the variation of the
initial amplitude of the Alfv\'en wave.

We also investigated the parametric space with regard to the
relative density perturbation. Namely,
we investigate dependence of $max(|(\rho-\rho_0)/ \rho_0|)$
as function of $\lambda$, $\beta$ and $A$.
We gather from Fig.~9D that for $\beta=2$ the generated
relative density perturbation saturates at about 20 \%,
while for $\beta=0.5$ the saturation level doubles.
Fig.~9E illustrates the fact that the maximum generated
relative density perturbation depends quite strongly on
plasma $\beta$.
In Fig.~9F we show the dependence of $max(|(\rho-\rho_0)/ \rho_0|)$
versus the initial amplitude $A$. We observe that since the density
perturbation is generated by the non-linear effects,
$max(|(\rho-\rho_0)/ \rho_0|)$ indeed grows with the increase of $A$,
and the dependence is almost linear.

\section{Conclusions}

This study is an extension of the
previous works to the case of 
strongly nonlinear amplitudes. The main results 
can be summarized as follows:
\begin{itemize}
\item  Phase mixing of a strongly nonlinear Alfv\'en pulse
is accompanied by an enhanced generation of compressible waves.
This is true irrespective of 
plasma $\beta$ being less or greater than unity.
\item  Plasma density
inhomogeneity (while providing, as in the weakly nonlinear case, a
source region for the non-linear generation of transverse
compressive waves by a plane Alfv\'en wave) substantially enhances
(by about a factor of 2) the generation of longitudinal
compressive waves. 
\item Attained maximal values of the
generated transverse compressive perturbations are  insensitive to
the strength of the plasma density inhomogeneity, plasma $\beta$ 
and the initial amplitude of the Alfv\'en wave.
Typically, they reach about 40\% of the initial Alfv\'en 
wave amplitude. 
\item Attained
maximal values of the generated relative density perturbations are
within  20-40\% for $0.5 \leq \beta \leq 2.0$. They depend upon
plasma $\beta$ strongly; and scale (increase) almost linearly
with the initial Alfv\'en wave amplitude. 
\end{itemize}

Commenting upon the maximal values attained by the
transverse compressive waves in relation to the previous,
weakly nonlinear, results \citep{Botha,td1} we would
like to state the following:
In the weakly nonlinear case the transverse compressive
waves do not grow to a substantial fraction of the initial
Alfv\'en wave amplitude 
due to the destructive wave interference effect
(typically, $max(|V_x(x,z,t)|)/A \sim 10^{-3}$).
However, in the strongly nonlinear case, studied here, 
although the destructive wave interference is still in action,
$max(|V_x(x,z,t)|)/A$ is now significantly larger, 
about $0.4$ (cf. Figs. 9A, 9B, 9C). 
Therefore, we conclude that the strong nonlinearity substantially
enhances the attained maximal values of the
transverse compressive waves.

The physical phenomenon studied here is an elementary process
responsible for non-resonant coupling of compressible and
incompressible MHD modes. In particular, it may play a role in MHD
turbulence of space and astrophysical plasmas. The presence of a
pressure-balanced inhomogeneity should significantly affect the
saturated MHD turbulent state (c.f. a similar claim but based upon
different reasonings in \citet{bns98}).

\acknowledgements D.T. acknowledges financial support from PPARC.
Numerical calculations of this work were
performed using the PPARC funded Compaq MHD Cluster at St Andrews
and Astro-Sun cluster at Warwick.

\end{document}